\documentclass[pre,twocolumn,
 amsmath,amssymb,
 aps,superscriptaddress
]{revtex4-2}

\usepackage{adjustbox}
\usepackage{graphicx}
\usepackage[toc,page]{appendix}
\usepackage[normalem]{ulem}
\usepackage[dvipsnames]{xcolor}
\usepackage{graphicx}
\usepackage{dcolumn}
\usepackage{bm}
\usepackage{float}
\usepackage{colortbl}
\usepackage{braket}

\newcommand{\beq}{\begin{equation}}
\newcommand{\eeq}{\end{equation}}

\begin{document}

\preprint{APS/123-QED}

\title{A computational approach to investigate TDP-43 C-terminal fragments aggregation in  Amyotrophic Lateral Sclerosis}

\author{Greta Grassmann}

\affiliation{Universit\`a di Bologna, Via Irnerio 46, Bologna I-40126, Italy.}

\author{Mattia Miotto} 
\affiliation{Center for Life Nano-\& Neuro-Science, Istituto Italiano di Tecnologia, Viale Regina Elena 291, 00161 Rome, Italy.}

\author{Lorenzo Di Rienzo}
\affiliation{Center for Life Nano-\& Neuro-Science, Istituto Italiano di Tecnologia, Viale Regina Elena 291, 00161 Rome, Italy.}

\author{Federico Salaris}

\affiliation{Center for Life Nano-\& Neuro-Science, Istituto Italiano di Tecnologia, Viale Regina Elena 291, 00161 Rome, Italy.}

\author{Beatrice Silvestri}
\affiliation{Department of Biology and Biotechnologies ``Charles Darwin", Sapienza University of Rome, Piazzale Aldo Moro 5, 00185 Rome, Italy.}
\affiliation{Center for Life Nano-\& Neuro-Science, Istituto Italiano di Tecnologia, Viale Regina Elena 291, 00161 Rome, Italy.}

\author{Elsa Zacco}
\affiliation{Department of Neuroscience and Brain Technologies, Istituto Italiano di Tecnologia, Via Morego 30, 16163 Genoa, Italy.}

\author{Alessandro Rosa}
\affiliation{Department of Biology and Biotechnologies ``Charles Darwin", Sapienza University of Rome, Piazzale Aldo Moro 5, 00185 Rome, Italy.}
\affiliation{Center for Life Nano-\& Neuro-Science, Istituto Italiano di Tecnologia, Viale Regina Elena 291, 00161 Rome, Italy.}

\author{Gian Gaetano Tartaglia}
\affiliation{Department of Neuroscience and Brain Technologies, Istituto Italiano di Tecnologia, Via Morego 30, 16163 Genoa, Italy.}
\affiliation{Center for Life Nano-\& Neuro-Science, Istituto Italiano di Tecnologia, Viale Regina Elena 291, 00161 Rome, Italy.}
\affiliation{Centre for Human Technologies, via Enrico Melen, 83, 16152 Genova, Italy.}

\author{Giancarlo Ruocco}
\affiliation{Center for Life Nano-\& Neuro-Science, Istituto Italiano di Tecnologia, Viale Regina Elena 291, 00161 Rome, Italy.}
\affiliation{Department of Physics, Sapienza University of Rome, Piazzale Aldo Moro 5, 00185 Rome, Italy.}

\author{Edoardo Milanetti}
\affiliation{Department of Physics, Sapienza University of Rome, Piazzale Aldo Moro 5, 00185 Rome, Italy.}

\affiliation{Center for Life Nano-\& Neuro-Science, Istituto Italiano di Tecnologia, Viale Regina Elena 291, 00161 Rome, Italy.}

\begin{abstract}
Many of the molecular mechanisms underlying the pathological aggregation of proteins observed in   neurodegenerative diseases are still not fully understood. Among the diseases associated with protein aggregates, for example, Amyotrophic Lateral Sclerosis (ALS) is of relevant importance.
Although understanding the processes that cause the disease is still an open challenge, its relationship with protein aggregation is widely known. In particular, human TDP-43, an RNA/DNA binding protein, is a major component of pathological cytoplasmic inclusions
described in ALS patients.
The deposition of the phosphorylated full-length TDP-43 in spinal cord cells has been widely studied, and it has been shown that the brain cortex presents an accumulation of phosphorylated C-terminal fragments (CTFs). 
Even if it is debated whether CTFs represent a primary cause of ALS, they are a hallmark of TDP-43 related neurodegeneration in the brain. Here, we investigate the CTFs aggregation process, providing a possible computational model of interaction based on the evaluation of shape complementarity at the interfaces. To this end, extensive Molecular Dynamics (MD) simulations were conducted for different types of fragments with the aim of exploring the equilibrium configurations. Adopting a newly developed approach based on Zernike polynomials, for finding complementary regions of the molecular surface, we sampled a large set of exposed portions of the molecular surface of CTFs structures as obtained from MD simulations. The analysis proposes a set of possible associations between the CTFs, which could drive the aggregation process of the CTFs.
\newline
\emph{keywords:} Molecular Dynamics simulation, Protein aggregation, Binding regions, TDP-43, Amyotrophic Lateral Sclerosis.
\end{abstract}
\maketitle

\section{Introduction}
The investigation of the molecular mechanisms that lead to the accumulations of aggregated proteins is  crucial  for understanding the pathophysiology of many neurodegenerative diseases \cite{baloh2011tdp}. The accumulation of aggregates containing the DNA- and RNA-binding protein TDP-43  in the central nervous system is a common feature in diseases, such as Amyotrophic Lateral Sclerosis (ALS), Frontotemporal Dementia (FTD), and Alzheimer’s Disease (AD), t \cite{zuo2021tdp, jo2020role}. However, the mechanisms of aggregation  are not yet fully understood and various aggregation models have been proposed \cite{baralle2013role}.
In this scenario, the fundamental role of the C-terminal fragments in the formation of TDP-43 aggregates has already been widely confirmed \cite{Jiang2017, Lee2014, Afroz2017, Wang2013, Tavella2018, Prasad2019}. In particular, it is interesting to study this aggregation model for the possible implications on ALS disease.\\
TDP-43 is composed of an N-terminal domain (NTD), two RNA recognition motifs (RRMs), and a long C-terminal (CTD) glycine-rich region \cite{jiang2017n}. 
The CTF is formed by  the last 195-206 residues \cite{Wang2013} of TDP-43, corresponding to a portion of the full protein including only the CTD together with a truncated RRM2 fragment.
In the CTFs two kinds of RRM2 fragments can usually be found \cite{Guenther2018}: one corresponding to the cleavage at the residue 208 and the other corresponding to the cleavage at residue 219.

\begin{figure*}[t]
\centering
\includegraphics[width=\textwidth]{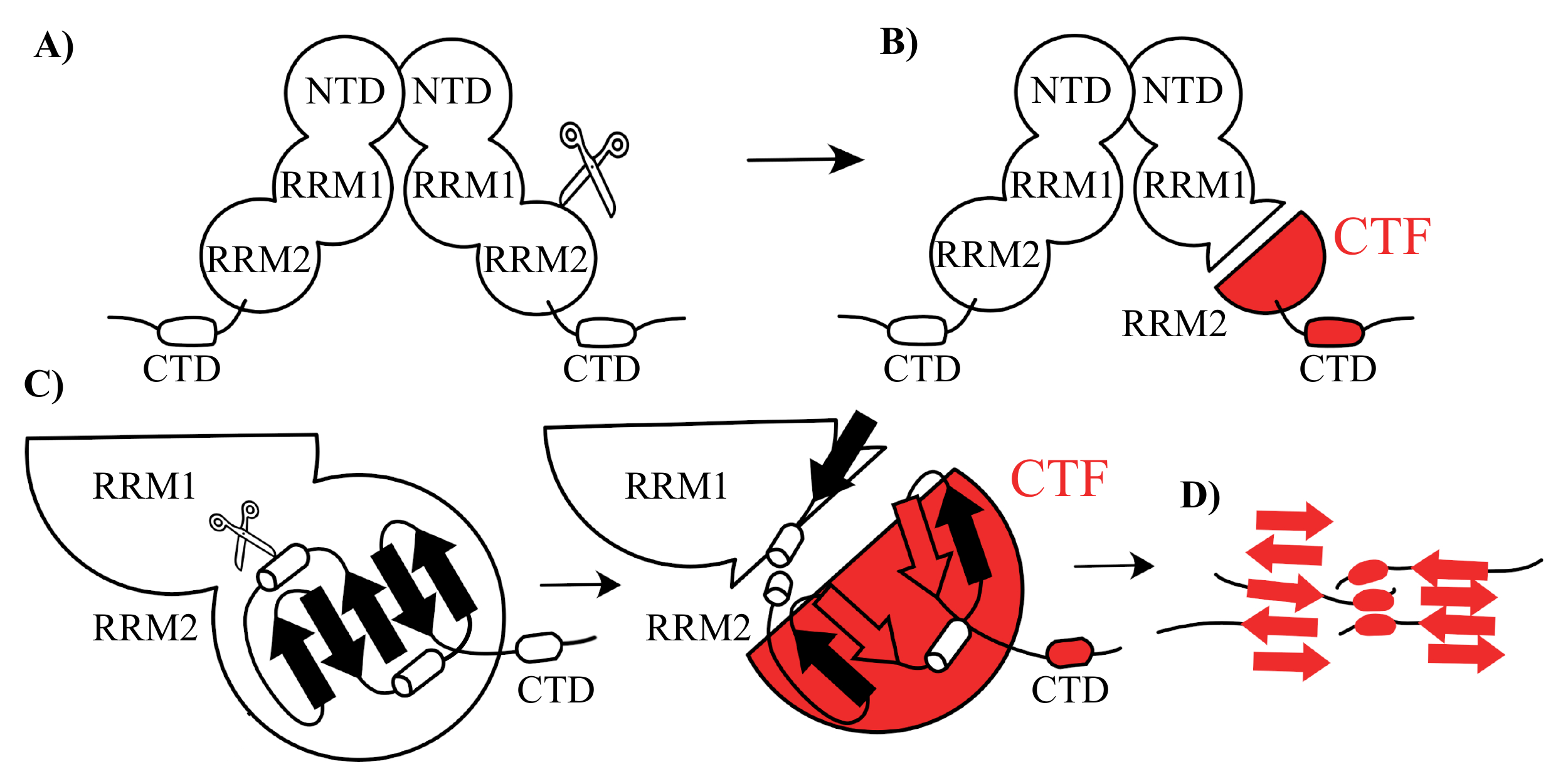}
\caption{ \textbf{Hypothesised model of the TDP-43 CTFs aggregation.}
\textbf{A)} TDP-43 in physiological conditions forms dimers. \textbf{B)} After the cleavage the CTF is split from the whole protein.
\textbf{C)} The RRM2 fragment resulting from the cleavage exposes its $\beta$-strands. \textbf{D)} The $\beta$-strands from different CTFs allow the formation of aggregates to happen.  }
\label{tdp43}
\end{figure*}

\begin{figure*}[t]
\centering
\includegraphics[width=.9\textwidth]{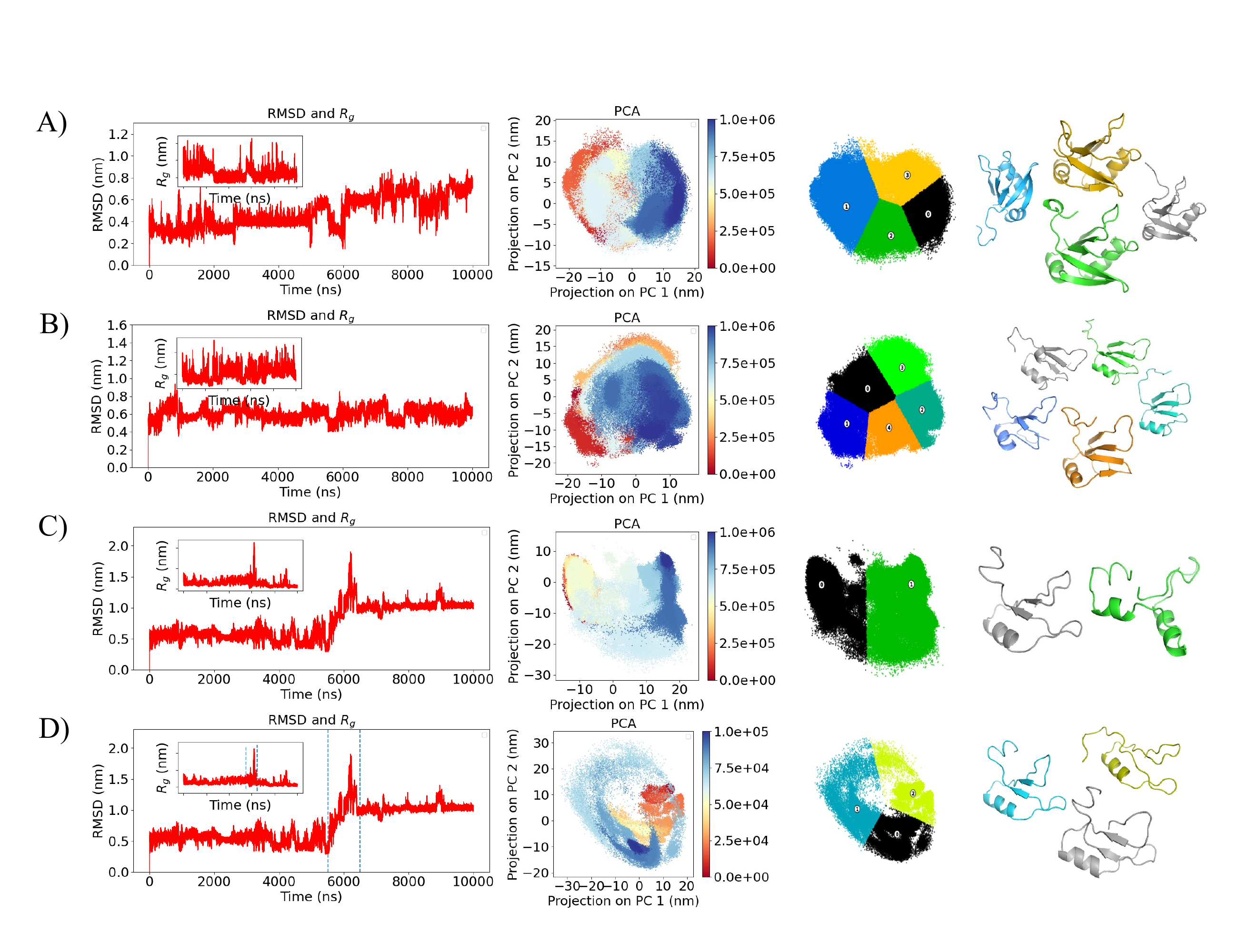}
\caption{ \textbf{Analysis of the Molecular Dynamics simulations.} \textbf{A)} From left to right, 
(i) Root Mean Square Deviation (RMSD) as a function of time for the whole RRM2 structure with respect to its equilibrated starting state. The Radius of gyration ($R_g$) during the simulation is shown in the inset.
(ii) Two-dimensional projection of the sampled conformations in the subspace spanned by the first two Principal Components (PCs) of the covariances of the atomic positions during the simulation. Each point corresponds to a conformation after a number of steps indicated by the color bar.
(iii) Clustering of the scatter plot of the two-dimensional projection of the sampled conformations. The centroids of each cluster (labeled by the numbered white circle) are depicted as well.(iv) Cartoon representation of the configurations corresponding to the found centroids.
\textbf{B)} Same as in A) but for Fragment A.
\textbf{C)} Same as in A) but for Fragment B.
\textbf{D)} Same as in C) but considering only the unfolding configurations of Fragment B, marked but the  blue vertical lines in RMSD plot.}
\label{conf}
\end{figure*}

\begin{figure}[t]
\includegraphics[width=0.75\columnwidth]{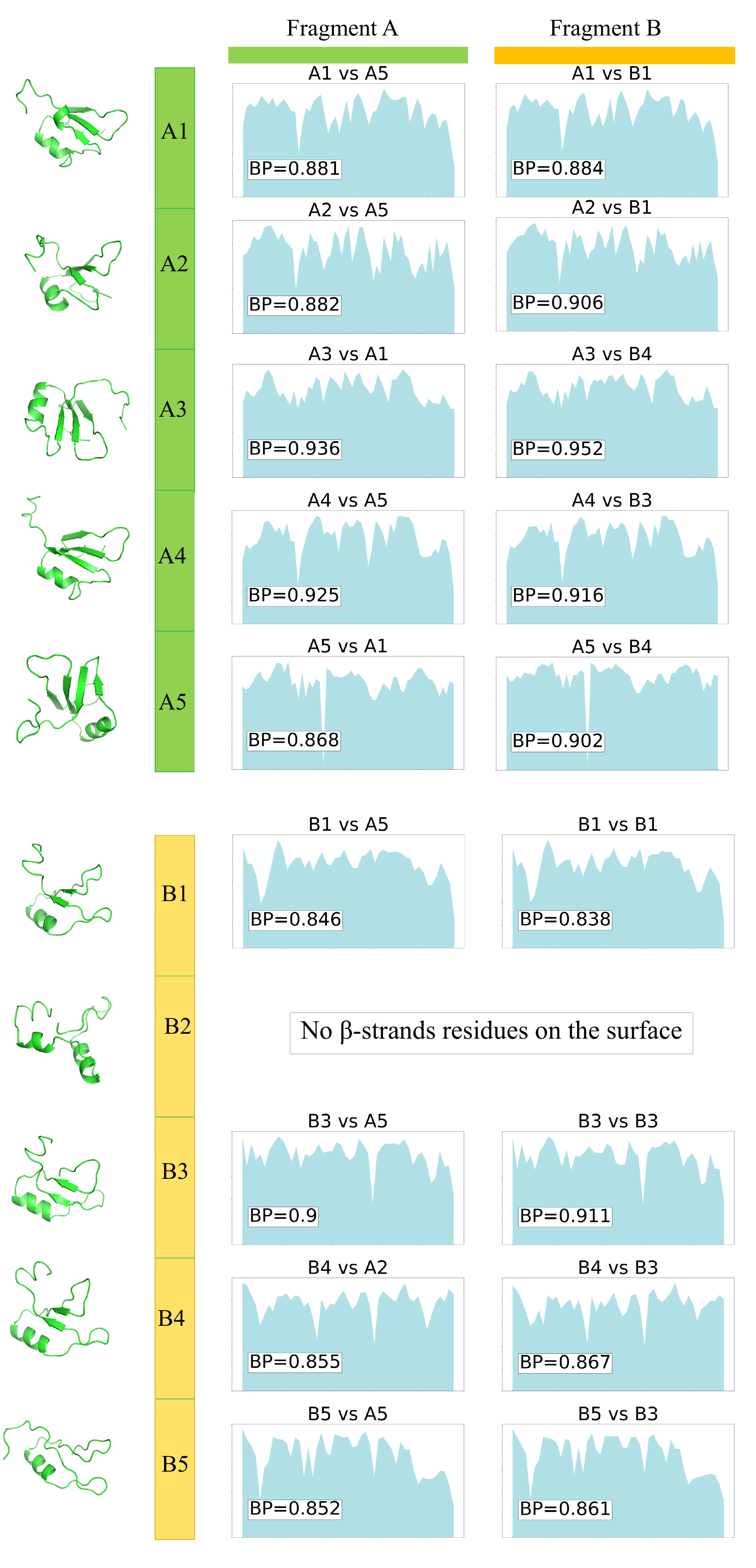}
\centering
\caption{\textbf{Binding propensity profiles.} Binding propensities scores for the conformations where the mean BP of the $\beta$-strand residues in the starting conformation has the highest value. Each row corresponds to one of the ten conformations: for each row $i$, the plot on the left corresponds to the BP of the residues of conformation $i$ respect to the conformation of Fragment B that results in the highest mean BP for the $\beta$-strands residues. The plot on the right instead shows the same results, but with the best pairing with a conformation of Fragment B. The row corresponding to B2 is empty because this conformation has no $\beta$-strand residues on its surface.}
\label{bp_res}
\end{figure}

In normal conditions, the NTD-driven head-to-tail oligomerization spatially separates the high aggregation-prone CTDs of consecutive TDP-43 monomers, antagonizing aggregation \cite{Afroz2017}.
However, if a preoteolytic cleavage releases the CTF, these free portions of the protein are able to aggregate \cite{Wang2013}: according to current knowledge, the formation of inclusions seems indeed to start from this disruption of the physiological oligomerization of TDP-43.
Furthermore, the removal of the N-terminus increases the cytoplasmic localization, since it deprives the CTF of the Nuclear Localization Signal (NLS). \\
The aggregation involves the CTD, which is intrinsically disordered and aggregation-prone, and harbors most of the mutations related to ALS \cite{Prasad2019}.
That said, it has been discussed how the CTD is necessary for cytoplasmic aggregation and toxicity but not sufficient, since it requires an intact RRM, i.e.  the RRM2 fragment in the CTFs is fundamental for the aggregation~\cite{Johnson2008}.
In physiological conditions RRM2 is a really stable domain, thanks to a cluster of twelve connected hydrophobic residues in its core \cite{Tavella2018}, but the cleavage deprives it of its stabilizing interaction with RRM1.
In a study of the RRM2 unfolding \cite{Mackness2014} that follows the separation from RRM1, it has been found that the mutually stabilizing interaction between RRM1 and RRM2 reduces the population of an intermediate state of the latter linked with pathological misfolding. This intermediate state may enhance the access to the Nuclear Export Signal (NES) within its sequence that increases the transport to the cytoplasm, and serve as a molecular hazard linking physiological folding with pathological misfolding and aggregation.\\
A second effect of the cleavage of RRM2 is the exposition of its aggregation-prone $\beta$-strands \cite{Wang2013, Kumar2019}. These strands are normally buried in the native state, but have been found to form fibrils \textit{in vitro} \cite{Tavella2018}.\\
These processes confirm a fundamental role of the fragment of RRM2 in the CTFs for the protein's aggregation \cite{Kumar2019}.
In particular, the $\beta$-strands could be at the core of the aggregation, because they are able to form steric zippers which then, following a typical atomic model for amyloid fibril structure formation, give rise to amyloid structures. \cite{Wang2013}.
Amyloid fibrils consist of packed $\beta$-sheets that run parallel to the fibril axes. Each $\beta$-sheet adheres to its neighboring sheet through the side chains that project roughly perpendicular to the fibril axis, toward the neighboring sheet. This interdigitation between the side chains of mating sheets is the so-called steric zipper.\\
In support of the hypothesis that this kind of structure is at the base of the CTFs aggregation, as shown in Figure \ref{tdp43}, it has been found that some regions of RRM2 can form different classes of steric zipper structures \cite{Prasad2019,Guenther2018} at the core of the formation of these amyloid fibrils \cite{Nelson2006}.

The aggregation of TDP-43 is strongly influenced by the interaction with DNA and RNA: RNA aptamers are able to interfere with the aggregation kinetic, as a function of their nucleotides composition, binding affinity, and length \cite{Zacco2019}.\\
A correctly designed RNA aptamer should then be able to hinder the aggregation.
This molecule should have an effect after binding to the RRM2 fragment: assuming the validity of the \textit{cross-$\beta$ spine} model for the CTFs aggregation, by binding an aptamer to the RRM2 site that forms the "spine" of the fibril, we could prevent another CTF to bind to that site. This is indeed an almost mandatory choice since the RRM2 fragment is the only part of the CTFs that we can control (with both MD simulations and the \textit{Zernike}  \cite{Milanetti2021} method) because of the CTD disordered structure.\\
The designing of this interfering molecule should obviously consider the binding compatibility to the misfolded conformation of the RRM2 fragments, which is not available yet in literature.\\
Because of this, we use MD simulations to study the conformations of the fragments after the cut. The aim of this work is  to suggest some possible binding regions that in the future could be taken as starting point for designing specific interfering molecules.\\
To obtain a complete study of this region of TDP-43, we employ MD simulations to study the evolution of the whole RRM2 as well.

\section{Results}
\subsection{Molecular dynamics simulations and equilibrium configurations}
A MD simulation of $10~\mu s$ is performed for the whole RRM2 and for each fragment  (see Methods section for details).\\
The whole RRM2 corresponds to the residues 192-269 of TDP-43.
The first fragment, Fragment A, corresponds to the residues 209-269. The second one, Fragment B, corresponds to the residues 220-269.\\
To obtain their equilibrium configurations, we firstly apply a Principal Component Analysis (PCA) on the trajectory resulting from each MD simulation. In this way, we get an essential representation of their motions.
Then, we implement a cluster analysis on the projection of each trajectory on its first two principal components. Our aim is to find the most representative configurations for each one of the possible conformations that the fragment can take at equilibrium: we are assuming that each cluster's center (or centroid) is a good representative of that cluster.\\
The structures corresponding to these centroids are the equilibrium configurations, we will use to search for possible binding regions  with the \textit{Zernike} polynomial formalism~\cite{Milanetti2021,latto,bo2020exploring}.
\\
We found four equilibrium configurations for the whole RRM2 (W1, W2, W3, W4), five for Fragment A (A1, A2, A3, A4, A5), and two for Fragment B (B1, B2).\\
Since the plot of the RMSD evolution shows a clear peak for the trajectory of Fragment B, which should correspond to an unfolding of the fragment, we implement the same analysis specifically for that time interval. After the application of the just-described procedure, we found as the most representative conformations for the unfolding of Fragment B three configurations: B3, B4, and B5.
Figure \ref{conf} shows the just described steps of the analysis and their results for each of the studied cases.

We then compute for each one of the conformations found for Fragment A and B the corresponding 3D molecular surface (see Methods for details). The shape complementarity between these molecular surfaces can now be studied with the \textit{Zernike} polynomial expansion.

\subsection{2D \textit{Zernike} polynomial expansion for binding regions prediction}
Recently, a new method based on the  \textit{Zernike} 2D polynomial expansion, has been developed with the aim of evaluating whether and where two proteins can interact with each other to form a complex, based on their shape complementarity~\cite{Milanetti2021}.
Although there is noise in the shape of the interacting molecular surface, which probably has not been evolved to optimize intermolecular binding, we have previously shown the goodness of the methods of recognizing interacting surface portions from non-interacting patches (decoys) \cite{Milanetti2021}. Not all complexes are characterized by the same degree of shape complementarity, but in general, this property is one of the most important for molecular docking. 
By expanding the well-exposed molecular surface patches in terms of 2D \textit{Zernike} polynomials, we are able to rapidly and quantitatively measure the geometrical complementarity between interacting proteins by comparing their molecular surfaces.\\
In our case, we apply it to all the possible pairs between the 3D surfaces of the two CTFs RRM2 fragments to find the binding regions on each surface.\\
More specifically, for each point $i$ belonging to the molecular surface we define a molecular surface region (patch), which can be described through 2D \textit{Zernike} formalism with a set of invariant descriptors. Two similar patches, if defined by the same reference point, have a small distance between the \textit{Zernike} vectors, since perfectly complementary patches are equal under roto-translation.
For each point, $i$ of one of the two surfaces, the distance between the \textit{Zernike} descriptors of its patch and all the patches built on the points of the other surface is computed. The minimum of these values is selected, and after all the points had been studied these minimum values are mapped in $[0,1]$ and inverted. At the end of the process, the points whose corresponding patches have a high complementarity with the other surface are associated with a value of the binding propensity (BP) near one.
As a next step, each point is associated with the mean BP value of the points in its neighborhood: the interacting regions should be made up mostly of elements with high complementarity and therefore a high average value of BP.\\
Finally, we associate to each residue the mean value of the BP of the corresponding points in the surface.
We apply this procedure to all the possible pairs of conformations. To identify the most promising regions of interaction between the two fragments (i.e., the regions that we expect to be at the core of the CTFs aggregation), we select, for each conformation of each fragment, the pairing, for each fragment, that results in the highest mean BP of the residues corresponding to $\beta$-strands in the conformation sequence. In this way, we are selecting the pairings that are more prone to bind through $\beta$ strands.\\
Figure \ref{bp_res} shows the results of this selection employed for each of the ten conformations for Fragment A and B.

\subsection{Proposal of binding and $\beta$-strands regions }
The pairings found in the previous section are relevant to describe  CTFs aggregation process.\\
Artificial molecules, such as RNA aptamers or peptides could be in the future proposed as candidates for the interruption of the molecular interaction between the CTFs of the TDP43 protein: to propose some binding regions suited for testing with aptamers we followed a second approach.\\
For each conformation, we summed the BPs obtained for all of its possible pairings with the other conformation.
In this way, we obtain a clear representation of the residues in each conformation that are in general more involved in the interaction with other surfaces.
At this point, we have to take into account the fact that the \textit{Zernike} method looks only at the shape complementarity between surfaces, which is a necessary but not sufficient condition for interaction. 
To select among the residues found with this summation, the ones corresponding to the binding regions that will be able to bind an aptamer, we can apply some additional chemical-based constraint.\\
In particular, we consider one of the most basic requirements for interaction, i.e. the Coulombic interaction: aptamers are characterized by a negative charge, consequently we can select among the \textit{Zernike}-selected residues the ones associated with a positive charge.

\begin{figure}[t]
\includegraphics[width=\columnwidth]{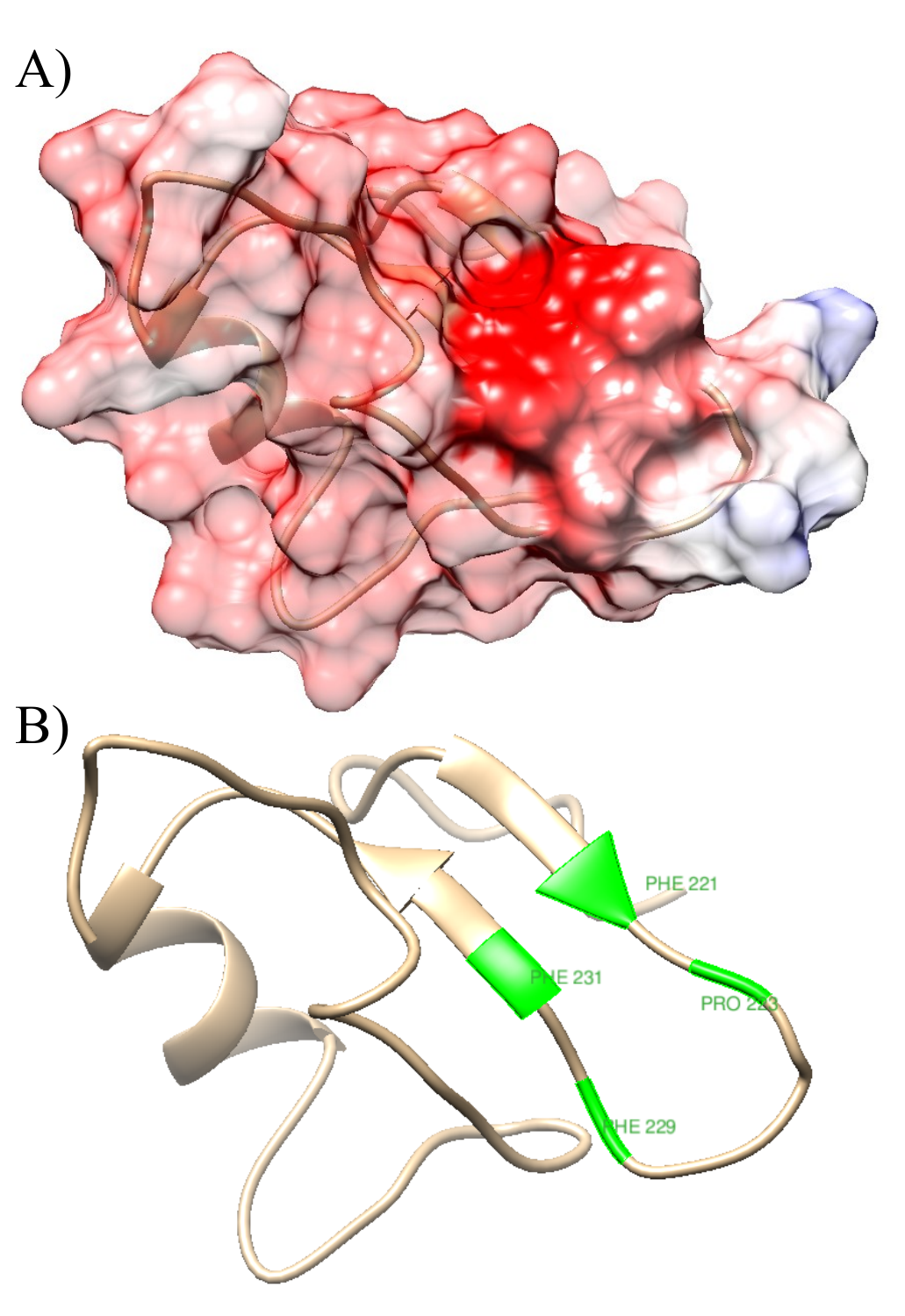}
\centering
\caption{\textbf{Example of binding region selection.}  
\textbf{A)} Coulomb surface colouring of the binding region of the first binding region for conformation A1.
\textbf{B)} Selection of the binding region sequence.}
\label{visual}
\end{figure}

To select these regions we use \textit{UCSF Chimera} \cite{Pettersen2004} to visualize the Coulombic surface coloring of each conformation: negative regions are red-colored, positive ones are blue-colored.
Then we look at the Coulomb coloring of the surface regions corresponding to our \textit{Zernike}-selected residues and select only the residues corresponding to non-negative regions.
Figure \ref{visual} shows an example of how a binding region is selected.

That said, we are interested in the study of the aggregation of these fragments as hypothesized by a model according to which their interaction is mediated by the $\beta$-strands. To better understand the importance of these $\beta$-strands we select, among the \textit{Zernike}-found residues, the ones corresponding to $\beta$-strand fragments as well, independently of their charge.\\
These processes result in the selection of the binding and $\beta$-strand regions listed in Table \ref{bp}. The residues among the binding regions colored in red are the ones corresponding to $\beta$-strand fragments. We expect these regions to be the most interesting ones since they are regions that according to our hypothesized model are at the core of the aggregation (because they include $\beta$-strands) and we will be able to test them by binding to them an aptamer (since they are positive).

\section{Discussion}
The principal objective of the here presented work was to find some regions on the TDP-43 CTFs (in particular on their RRM2 fragment) to propose as the foundation of their aggregation.\\
Since the structure of these fragments had not been studied yet and their conformations were not yet available, we began our project by studying the time evolution of the two possible RRM2 fragments constituting the CTFs, i.e. Fragment A and B, with MD simulations; we studied the whole RRM2 as well.\\
From the analysis of the trajectories of these two fragments, we found five representative configurations each: these are the possible configurations that should be observable in a cell, as well as the ones that the fragments assume while interacting with each other. The definition of this equilibrium configuration is the first result presented by this paper.\\
As a next step, we searched on the surfaces of these configurations all the possible regions of interaction, by verifying their shape complementarity. This complementarity is a necessary but not sufficient condition for interaction: even though having narrowed the possible region of interaction to a limited set of residues (listed in Table \ref{bp}) is already an interesting result, this study could be in the future further developed with additional constraints for the selection of regions.\\
A second continuation of this work could be the verification \textit{in vitro} of our results: following the insertion of expressively designed aptamers (starting from our suggested binding regions' residues) in CTFs-expressing cells, if our results are correct, the aggregates number and dimension should decrease.
With this aim, we identified among the \textit{Zernike} selected binding regions the ones that would be able to bind an aptamer.

\begin{table}[t]
\begin{center}
\begin{adjustbox}{max width=\columnwidth}
\begin{tabular}{ |c||c|  }
 \hline
 \rowcolor{lightgray}\multicolumn{2}{|c|}{Conformation A1} \\
 \hline \hline
 I binding region &  \textcolor{red}{PHE221}, PRO223, PHE229, \textcolor{red}{PHE231}\\
 \hline
  II binding region & 	GLN213, TYR214, ILE250, LYS251 \\
 \hline
 \hline
 I $\beta$-strand region & 	PHE221\\
 \hline
II $\beta$-strand region & 	PHE231 \\
 \hline
 %%%%%  A2
 \hline
 \rowcolor{lightgray}\multicolumn{2}{|c|}{Conformation A2} \\
 \hline \hline
 I binding region & SER212, \textcolor{red}{GLN213}, \textcolor{red}{TYR214}\\
 \hline
 \hline
 I $\beta$-strand region & 	GLN213, TYR214, GLY215, ASP216, VAL217, \\
 & MET218, ASP219, VAL220, ILE222\\
 \hline
 %%%%       A3
  \hline
 \rowcolor{lightgray}\multicolumn{2}{|c|}{Conformation A3} \\
 \hline \hline
 I binding region & PRO223, ARG227, PHE229, PHE231\\
 \hline
 \hline
 I $\beta$-strand region & 	SER254, VAL255, IHS256\\
 \hline
 %%%%%%%%   A4
   \hline
 \rowcolor{lightgray}\multicolumn{2}{|c|}{Conformation A4} \\
 \hline \hline
 I binding region & \textcolor{red}{PHE221}, PHE226, ARG227, \textcolor{red}{ALA228}, \textcolor{red}{PHE229}, \textcolor{red}{PHE231} \\
 \hline
 II binding region & \textcolor{red}{VAL255}, \textcolor{red}{HIS256}, \textcolor{red}{ILE257}, \textcolor{red}{SER258} \\
 \hline
 \hline
 I $\beta$-strand region & 	PHE221, ALA228, PHE229, PHE231, THR233 \\
 \hline
 II $\beta$-strand region & SER254, VAL255, HIS256, ILE257, SER258\\
 \hline
 %%%%%%%    A5
  \hline
 \rowcolor{lightgray}\multicolumn{2}{|c|}{Conformation A5} \\
 \hline \hline
 I binding region & \textcolor{red}{PHE221}, \textcolor{red}{ILE222}\\
 \hline
 \hline
 I $\beta$-strand region & VAL217, MET218, ASP219, VAL220, PHE221, ILE222 	\\
 \hline
 II $\beta$-strand region & THR233, PHE234 \\
 \hline
 III $\beta$-strand region & ILE257 \\
 \hline
  \hline
 %%%%%% B1
 \rowcolor{lightgray}\multicolumn{2}{|c|}{Conformation B1} \\
 \hline \hline
 I binding region &	VAL220, ILE222, LEU248, ILE250 \\
 \hline
  II binding region & 	ARG227, ALA228, \textcolor{red}{PHE229}	 \\
 \hline
 \hline
 I $\beta$-strand region &	PHE229 \\
 \hline
II $\beta$-strand region & 	IHS256, ILE257, SER258 \\
 \hline
 %%%%%  B2
 \hline
 \rowcolor{lightgray}\multicolumn{2}{|c|}{Conformation B2} \\
 \hline \hline
 I binding region & LYS224, PRO225, PHE226, ARG227, ALA228, PHE231,\\ 
 & VAL232, THR233, PHE234, ALA235, ILE239 \\
 \hline
 II binding region &	HIS256, ILE257 \\
 \hline
  \hline
 %%%%%% B3
 \rowcolor{lightgray}\multicolumn{2}{|c|}{Conformation B3} \\
 \hline \hline
 I binding region &	VAL220, ILE222, LYS224, LEU248, ILE250	 \\
 \hline
  II binding region & 	ARG227, \textcolor{red}{PHE229}	 \\
  \hline
  III binding region &  ILE253\\
 \hline
 \hline
 I $\beta$-strand region & 	PHE229	 \\
 \hline
II $\beta$-strand region & 	HIS256, ILE257	 \\
 \hline
 %%%%%  B4
 \hline
 \rowcolor{lightgray}\multicolumn{2}{|c|}{Conformation B4} \\
 \hline \hline
 I binding region & VAL220, PHE221, ILE222, \textcolor{red}{PHE229}, \textcolor{red}{PHE231}, ILE249,\\ 
 & ILE250, GLY252, ILE253, \textcolor{red}{HIS256}, ASN267, ARG268 \\
 \hline
 \hline
 I $\beta$-strand region & 	PHE229, PHE231  \\
 \hline
 II $\beta$-strand region &	HIS256, ILE257, SER258 \\
 \hline
 %%%%%%%% B5
  \hline
 \rowcolor{lightgray}\multicolumn{2}{|c|}{Conformation B5} \\
 \hline \hline
 I binding region & VAL220, PHE221, ILE222, ARG227, \textcolor{red}{PHE229},\\ &\textcolor{red}{PHE231}, THR233, PHE234, ILE250, GLY252 \\
 \hline
 \hline
 I $\beta$-strand region & 	PHE229, PHE231  \\
 \hline
\end{tabular}
\end{adjustbox}
\end{center}
 \caption{Binding and $\beta$-strands regions found for each conformation.}
\label{bp}
\end{table}

\section{Materials and Methods}\label{methods}

\subsection{Dataset}
The three starting structures for the MD simulations of the whole RRM2, Fragment A, and Fragment B were extracted from the Protein Data Bank \cite{BERNSTEIN1977}: we begun from the Nuclear Magnetic Resonance (NMR) structure of the TDP-43 tandem RRMs in complex with UG-rich RNA (PDB id: 4BS2) \cite{Lukavsky2013}, and then deleted from the file the RNA e the RRM1 domain. The resulting  structure, to which we refers as `whole RRM2'  contains residues from 192 to 269 of TDP-43. 
Next, to obtain the molecular structure of Fragment A  removing  residues up to the 208$th$, whereas to obtain Fragment B we removed the residues up to the 219$th$.

\subsection{Molecular Dynamics simulations}
For each of the starting structures, we carried out one molecular dynamics simulation of 10 $\mu s$.\\  
All steps of the simulation were performed using Gromacs 2019.3~\cite{https://doi.org/10.5281/zenodo.3562495}.\\
The topologies of the system were built using the CHARMM-27 force field~\cite{charmm}, the standard force field for proteins.
Each fragment was placed in a rhombic dodecahedron simulative box, with periodic boundary conditions, filled with TIP3P water molecules~\cite{Jorgensen1983}. The system of the whole RRM2 included 5269 water molecules, Fragment A 4607 and Fragment B 4658. 
The rhombic dodecahedron box is built so that each atom of each fragment is at least at a distance of 11 \AA$~$from the box borders. Its volume is 71\% of the one of a cubic box of the same periodic distance: fewer water molecules have to be added to solvate the protein. For a protein to have the correct behavior there need to be at least two or three layers of water around it: with 11 \AA$~$there is space for approximately five layers.
The final system of the whole RRM2, consisting of 17038 atoms, was first minimized with 371 steps of steepest descent.
In the same way, the system of Fragment A, consisting of 14777 atoms, was minimized with 102 steps, whereas the system of Fragment B, consisting of 14759 atoms, was minimized with 346 steps. Each step had a size of $0.01$, while the force limit value was set to $max(|\textbf{F}_n|)<10^3~kJ/mol/nm$.\\
The thermalization and pressurization of the systems in NVT and NPT environments were run each for 0.1 $ns$ at 2 $fs$ time-step. The temperature was kept constant at 300 $K$ with a Modified Berendsen thermostat and the final pressure was fixed at 1 $bar$ with the Parrinello-Rahman algorithm \cite{parrinello} (with a time constant of coupling between the system and the barostat of $\tau_P=2~ps$), which guarantees a water density close to the experimental value of the SPC/E model of water of $1008~kg/m^3$. Indeed for the whole RRM2, we obtained an average density value of $1012\pm 4~kg/m^3$, for Fragment A a value of $1006\pm 5~kg/m^3$, and a value of $1002\pm 4~kg/m^3$ for Fragment B. LINCS algorithm \cite{lincs} was used to constraint h-bonds.\\
Finally, the systems were simulated with a 2 $fs$ time-step for 10 $\mu s$ in periodic boundary conditions, using a cut-off of 12 \AA$~$ for the evaluation of short-range non-bonded interactions and the Particle Mesh Ewald method \cite{Cheatham1995} for the long-range electrostatic interactions.\\
For all these steps the Leap-Frog integrator and the Verlet cut-off scheme were used.

\subsection{Computation of molecular surfaces}  
The molecular surfaces were obtained starting from the PDB files found after the clustering of the PCA of the trajectories resulting from the MD simulations. To compute for each one of the 14 structures the solvent-accessible surface, we used DMS \cite{richards1977areas}, with a density of 5 points per \AA$^2$ and a water probe radius of 1.4 \AA. For each surface point, we calculated the unit normal vector with the flag $-n$.

\subsection{Evaluation of Shape Complementarity}
The first step of this algorithm is to select from the surface a patch $\Sigma$, defined as
the set of surface points that fall within a sphere of radius $R_{zernike}=6$ \AA$~$ centered on one point of the surface.
The points contained in this sphere are divided, with a clustering from a random point that includes only the points closer than a distance $D_p$, in points belonging to the surface and points not directly connected to it (for example coming from a protuberance included in the sphere). Only the former will constitute the patch.\\
Once the patch has been selected, the mean vector of the normal vectors of the patch points is computed and oriented along the $z$-axis. Thus, given a point $C$ on the $z$-axis, we define $\theta$ as the largest angle between the $z$-axis and a
secant connecting $C$ to any point of the patch $\Sigma$. $C$ is then set so that $\theta=45^\circ$ and each surface point is labeled with its distance $r$ from $C$. As a next step, a square grid that associates each pixel with the mean $r$ value calculated on the points inside it is built.\\
Such a 2D function can now be expanded on the basis of the Zernike polynomials.
Indeed, each function of two variables $f(r,\psi)$ defined in polar coordinates inside the region of the unitary circle can be decomposed in the Zernike basis as
\begin{equation}
f(r,\psi)=\sum_{n'=0}^\infty\sum_{m=0}^{n'}c_{n'm}Z_{n'm}(r,\psi),
\end{equation}
with
\begin{equation}
    c_{n'm}=\frac{n'+1}{\pi}\int_0^1dr~r\int_0^{2\pi}d\psi Z_{n'm}^*(r,\psi)f(r,\psi)
\end{equation}
and 
\begin{equation}
    Z_{n'm}=R_{n'm}(r)e^{im\psi}.
\end{equation}
$c_{n'm}$ are the expansion coefficients, while the complex functions $Z_{n'm}(r,\psi)$ are the Zernike polynomials. The radial part $R_{n'm}$ is given by
\begin{equation}
    R_{n'm}(r)=\sum_{k=0}^{\frac{n'-m}{2}}\frac{(-1)^k(n'-k)!}{k!\big(\frac{n'+m}{2}-k\big)!\big(\frac{n'-m}{2}-k\big)!}.
\end{equation}
Since for each couple of polynomials, it is true that
\begin{equation}
   < Z_{n'm}|Z_{n''m'}>=\frac{\pi}{n'+1}\delta_{n'n''}\delta_{mm'},
\end{equation}
the complete sets of polynomials form a basis, and knowing the set of complex coefficients ${c_{n'm}}$ allows for a univocal reconstruction of the original patch.\\
Once a patch is represented in terms of its Zernike descriptors, the similarity between that patch and another one can be simply measured as
the Euclidean distance between the invariant vectors.\\
The norm of each coefficient $z_{n'm}=|c_{n'm}|$ constitutes one of the Zernike invariant descriptors. Since $z_{n'm}$ does not depend on the phase (i.e. it is invariant for rotations around the origin of the unitary circle), two patches can be assessed by comparing the Zernike invariants of their associated 2D projections, without considering their orientation. On the other hand, the relative orientation must be taken into account: if we search for similar regions we must compare patches that have the same orientation once projected in the 2D plane, i.e. the solvent-exposed part of the surface must be oriented in the same direction for both patches (for example as the positive z-axis).  If instead, we want to assess the complementarity between them, we must orient the patches contrariwise, i.e. one patch with the solvent-exposed part toward the positive z-axis (‘up’) and the other toward the negative z-axis (‘down’).
\\[14pt]
What is in practice done to understand if two surfaces have some complementary patches as described in the following:
\begin{enumerate}
    \item For both surfaces we compute the Zernike descriptors of the patches centered in all the points of the two surfaces up to the selected maximum expansion order $n$. The two surfaces have to be oriented in opposite verse along the $z$-axis: we are going to study the similarity between the first one and the inverse of the second one.
    \item For each point $i$ of the surface $1$, we compute the distance between the Zernike descriptors of its patch and all the patches built on the points of the surface $2$. The minimum of these values is selected, and after all the points have been studied these minimum values are mapped in $[0,1]$ and inverted. At the end of the process, the points whose corresponding patches have a high complementarity with the other surface are associated with a value near one.
    \item  After all surface points are associated with these binding propensities, we perform a smoothing process.\\
In this process, each point is associated with a novel binding propensity (BP) computed as the mean value of the points in its neighborhood, defined as all the points having a spatial distance from it smaller than 6 \AA.\\
\end{enumerate}

%\section*{Acknowledgment}

%\bibliography{mybiblio.bib}
\end{document}